\documentclass{PoS}
\usepackage{mathrsfs}
\usepackage{eucal}
\usepackage{lineno}

\title{Searching for neutrino emission from hard X-ray sources with IceCube}

\ShortTitle{Searching for neutrino emission from hard X-ray sources with IceCube}

\author{
The IceCube Collaboration\footnote{For collaboration list, see PoS(ICRC2019) 1177.}\\
{\itshape \href{http://icecube.wisc.edu/collaboration/authors/icrc19_icecube}{http://icecube.wisc.edu/collaboration/authors/icrc19\_icecube}}\\
E-mail: \email{santander@icecube.wisc.edu}
}

\abstract{

The IceCube neutrino observatory, a cubic-kilometer particle detector at the South Pole, first announced the discovery of an astrophysical flux of high-energy neutrinos in the TeV-PeV range in 2013, followed in 2017 by the detection of a high-energy neutrino event in temporal and directional correlation with the flaring gamma-ray blazar TXS 0506+056. This observation, combined with archival neutrino detections in 2014-2015, has provided compelling evidence for the detection of the first high-energy astrophysical neutrino source. 
A promising way of detecting additional sources is to correlate neutrino detections with sources where a hadronic electromagnetic signature is observed. If blazars are a significant source of neutrinos, the high-energy gamma rays produced in pionic decays in coincidence with the neutrinos may cascade in the strong photons fields present in blazar jets, leading to strong emission in the hard X-ray to MeV gamma-ray energy range. We here present plans for a search for neutrino emission from a large sample of hard X-ray sources from the BAT AGN Spectroscopic Survey (BASS).

\vspace{4mm}
{\bfseries Corresponding authors:}
{Marcos Santander}$^{1}$\\
{$^{1}$ \itshape Department of Physics and Astronomy, University of Alabama, Tuscaloosa, AL 35487-0324, USA}\\

}

\FullConference{36th International Cosmic Ray Conference -ICRC2019-\\
		July 24th - August 1st, 2019\\
		Madison, WI, U.S.A.}

\usepackage{amsmath}

\begin{document}

\section{Introduction}

High-energy neutrinos are unique tracers for hadronic processes occurring in astrophysical objects. Neutrinos in the TeV-PeV energy range are produced in the interactions of cosmic rays with ambient photon or matter fields at the source or during propagation. The charged pions produced in these interactions eventually decay into neutrinos which oscillate to flavor equipartition as they travel over cosmic distances. Their neutral charge and low interaction cross section implies that, unlike their parent cosmic ray particle, they can propagate in straight lines and suffer no absorption. The neutral pions produced in the same interactions decay into gamma rays in a similar energy range that accompany the neutrino emission~\cite{Kelner:2006tc, Kelner:2008ke}. 
These gamma rays can also point back to their source, but may be absorbed at the source or during propagation through interactions with the extragalactic background light (EBL)~\cite{2016RSOS....350555C}. 

\begin{figure}
    \centering
    \includegraphics[scale=0.53]{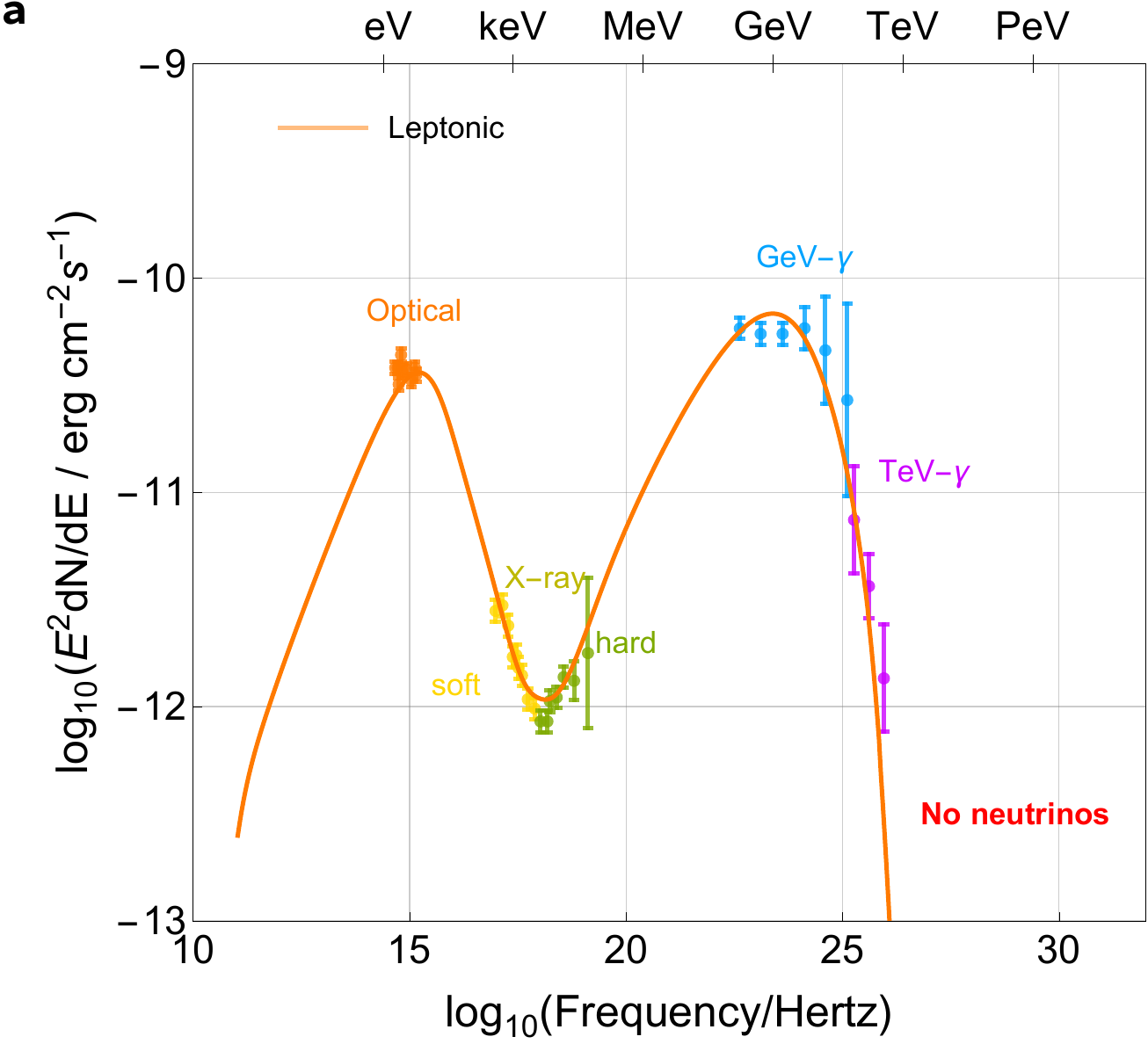}
    \includegraphics[scale=0.53]{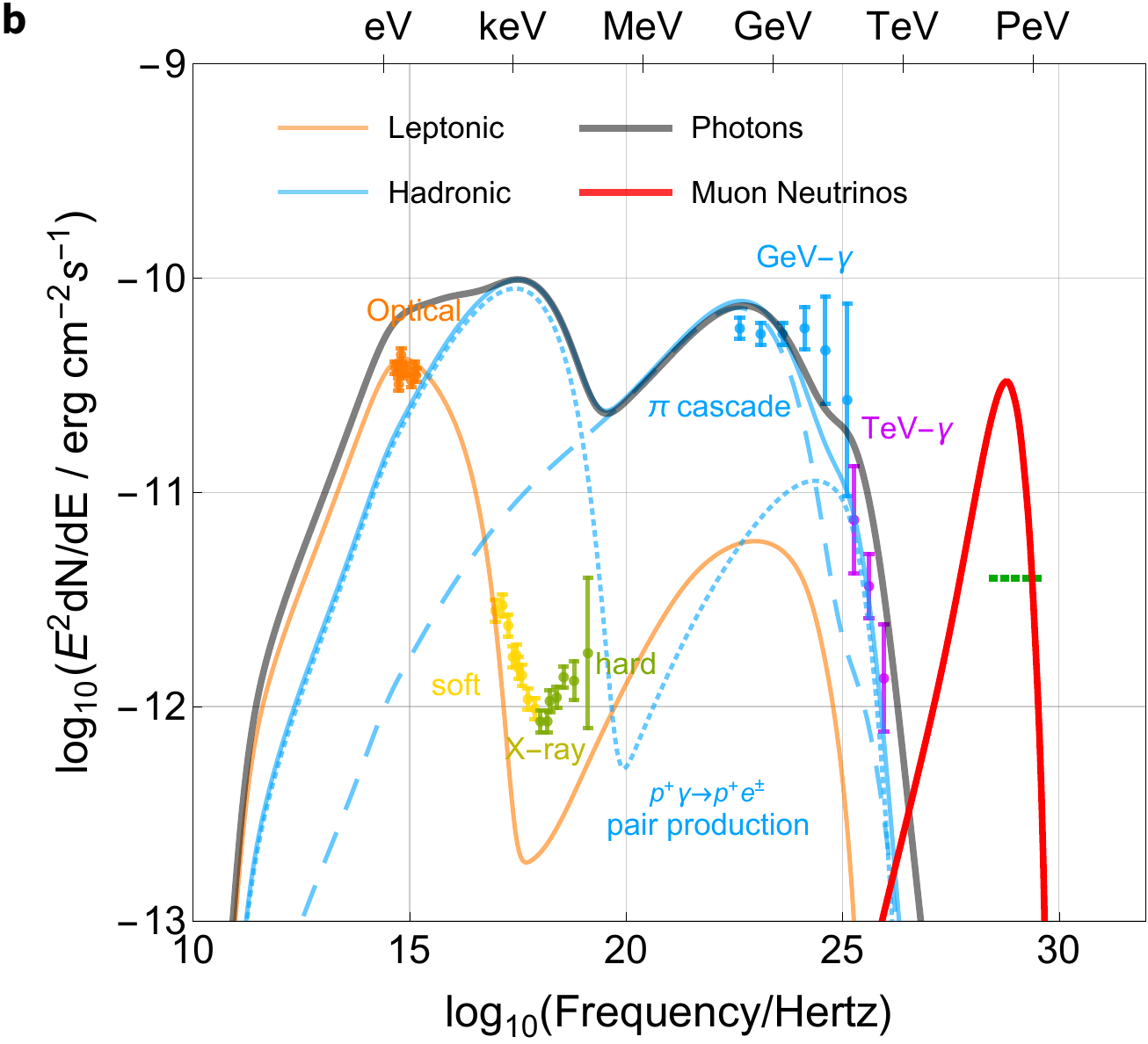}
    \caption{Modeling of the spectral energy distribution of the blazar TXS 0506+056 during the 2017 flare potentially associated with a high-energy neutrino for purely leptonic (\emph{a}) and purely hadronic (\emph{b}) models. The hadronic models tend to overproduce flux in the hard X-ray to MeV range due to hadronic pair production and cascading effects (from~\cite{Gao:2018mnu}).}
    \label{fig:txs}
\end{figure}

Hadronic gamma rays produced in photon-rich environments, such as Active Galactic Nuclei (AGN), may lose energy through electromagnetic cascading, which results in a high flux emitted in the hard X-ray to MeV range~\cite{Petropoulou:2015upa,PetroMast2015, Murase:2018iyl}. 
While these sources and the propagation medium may be opaque to TeV-PeV gamma rays, the X-ray to MeV emission may be observed largely unattenuated, and therefore represent an important probe for hadronic emission (see Fig.\ref{fig:txs} for an example using TXS 0506+056). 

The angular distribution of the astrophysical neutrinos discovered by IceCube in the TeV-PeV range~\cite{Aartsen:2013jdh} is consistent with isotropy, which favors an extragalactic origin for the signal. AGN are therefore a prime candidate neutrino source class, as they dominate the high-energy extragalactic sky. While no significant correlation has been observed between IceCube neutrinos and gamma-ray AGN from the \emph{Fermi} 2LAC catalog~\cite{Aartsen:2016lir, Huber}, the observation of a high-energy neutrino in 2017 in coincidence with a gamma-ray flare from the blazar TXS 0506+056 has provided the first evidence for neutrino emission from these objects~\cite{IceCube:2018dnn}. Further evidence for historical neutrino emission from TXS 0506+056 with no associated $\sim$GeV gamma-ray flare~\cite{IceCube:2018cha} points to a more complex scenario, although the lack of sensitive broad multiwavelength coverage, specially in the X-ray band, makes an interpretation challenging~(e.g.~\cite{Rodrigues:2018tku}).

While previous neutrino searches from AGN have concentrated on the correlation with gamma-ray emission, we here propose a search for neutrinos from a catalog of hard X-ray AGN detected with Burst Alert Telescope (BAT)~\cite{Barthelmy:2005hs} onboard the Neil Gehrels \emph{Swift} space telescope. 

\section{The hard X-ray AGN sample}

\begin{figure}
    \centering
    \includegraphics[scale=0.5]{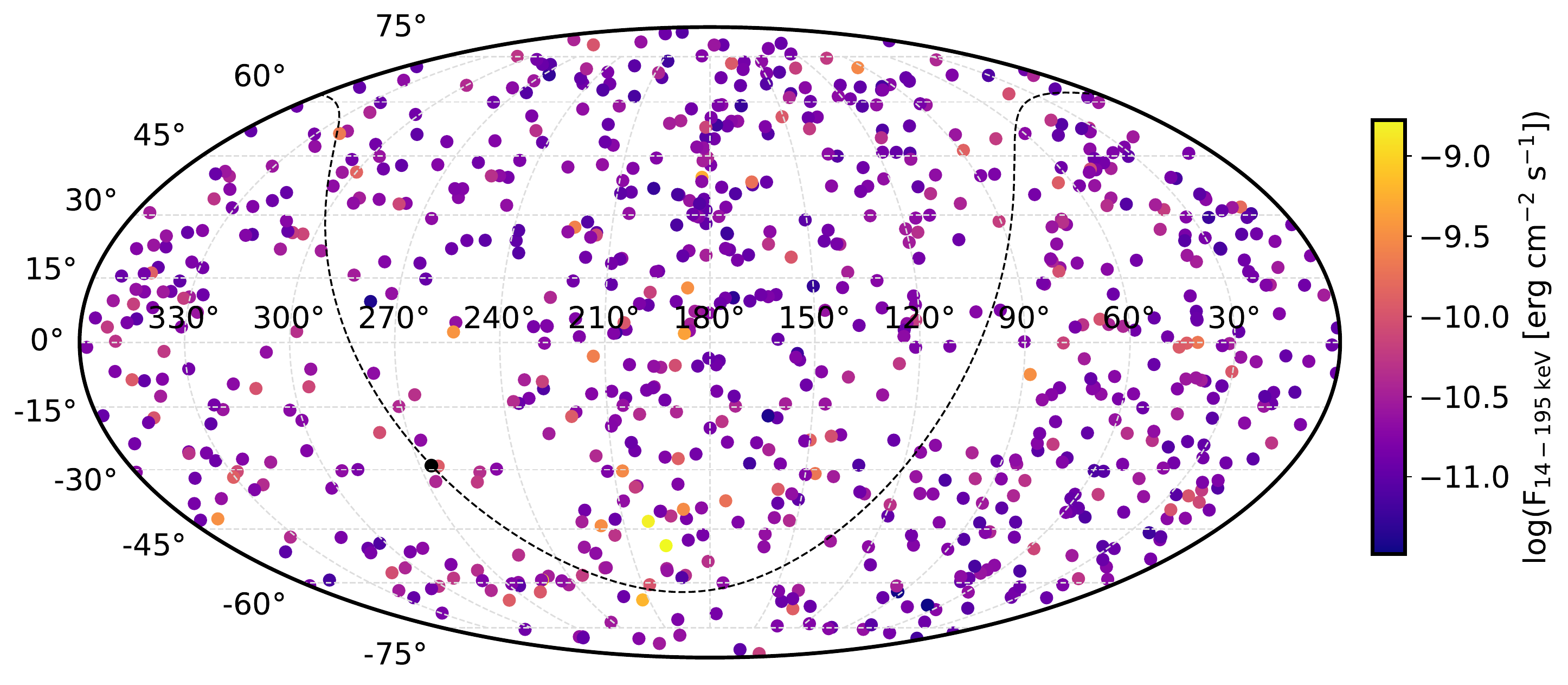}
    \caption{Skymap in equatorial coordinates of the BASS sources that will be used in the proposed search. The dashed line indicates the Galactic plane, with a black marker showing the Galactic Center position. Marker colors show the intrinsic (i.e. deabsorbed) energy flux of the source in the 14-195 keV band in logarithmic scale.}
    \label{fig:map}
\end{figure}

The first data release (DR-1) of the BAT AGN Spectroscopic Survey (BASS)\footnote{\href{https://www.bass-survey.com/}{https://www.bass-survey.com/}}~\cite{Koss:2017ilk} constitutes the most complete all-sky AGN catalog in the hard X-ray range (14-195 keV). The sources included in BASS were selected from the 70-month BAT catalog, which lists 1210 objects in the $>10$ keV range. Given the large positional uncertainty of BAT sources, \emph{Swift} XRT observations were used to identify counterparts which resulted in 836 BAT-detected AGN. Redshift values are listed for 828 AGN, which we use in this study. BASS includes dedicated and archival optical spectroscopic observations of the identified AGN for redshift determination and spectral line measurements. A skymap of all BASS sources used in this study is shown in Fig.~\ref{fig:map}.

Most BASS sources are nearby, with a median of $z\sim0.04$ and 90\% of the sources located within a redshift of 0.2. Approximately 11\% of the sources are potentially beamed (either BL Lac objects or quasars) where Doppler boosting increases the observed emission from the source.

\begin{figure}
    \centering
    \includegraphics[scale=0.4]{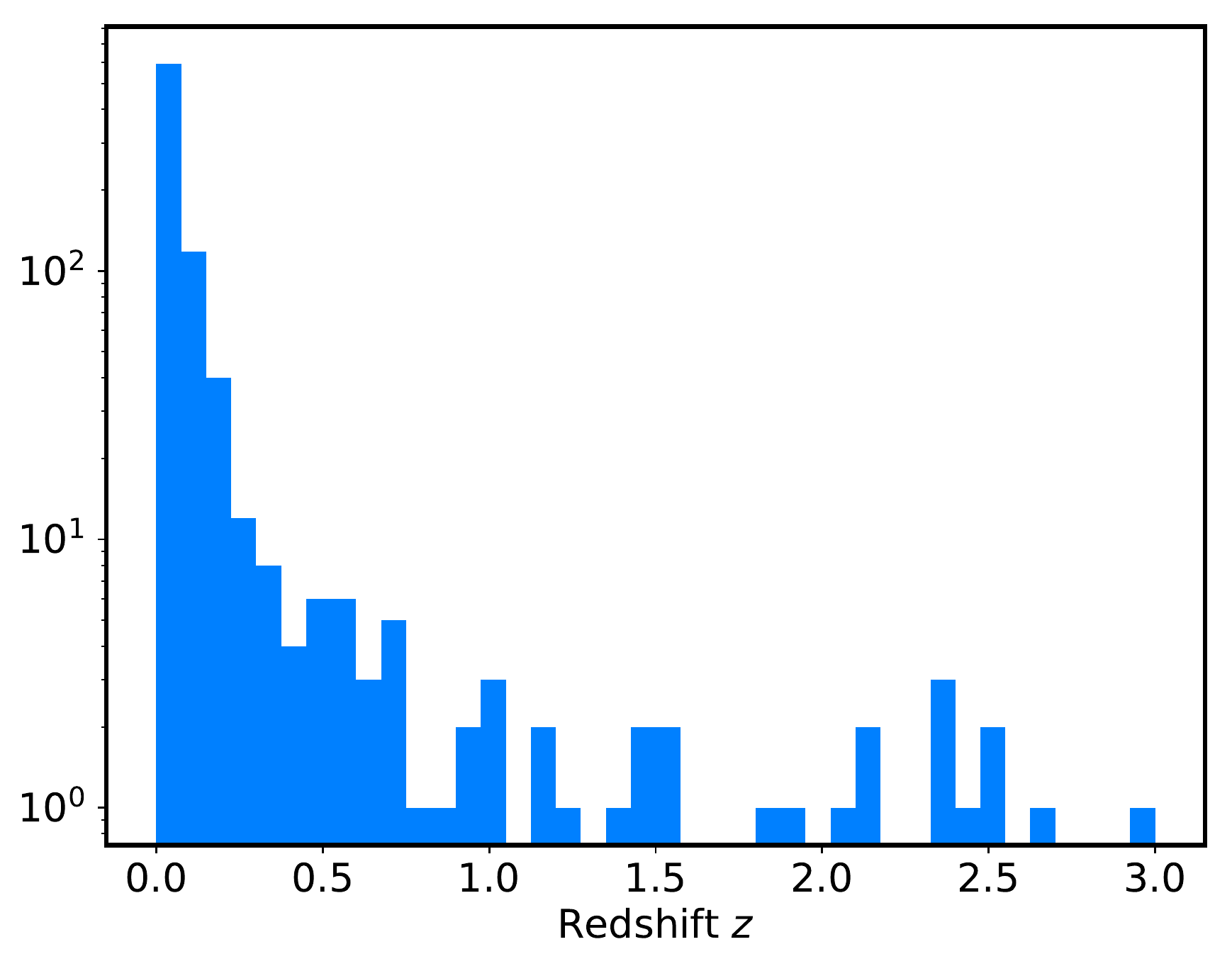}
    \includegraphics[scale=0.4]{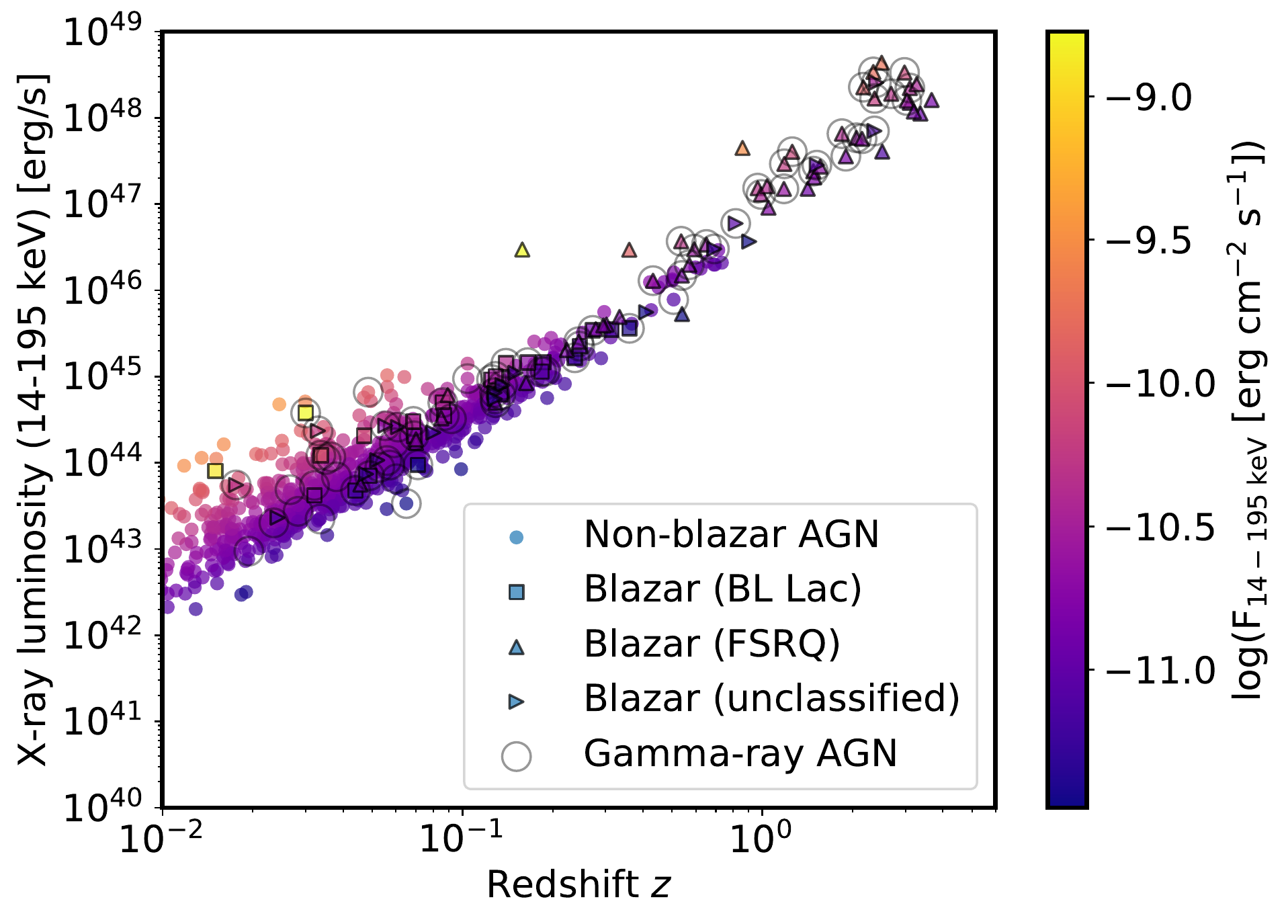}
    \caption{\emph{Left:} Redshift distribution of the BASS catalog sources. \emph{Right:} Hard X-ray luminosity (isotropic-equivalent) for the BASS sources as a function of redshift. Different markers indicate AGN classes, while the colors indicate the derived intrinsic luminosity in the 14-195 keV band. The highest luminosities are mostly associated with blazars thanks to the contribution of Doppler boosting. AGN with a potential gamma-ray counterpart from the \emph{Fermi}-LAT 4FGL catalog are marked with a gray circle.}
    \label{fig:zlumi}
\end{figure}

\subsection{Overlap with the \emph{Fermi} 4FGL catalog}

As several sources in BASS may also be listed in the \emph{Fermi} catalogs that have been used in the past to search for neutrino correlations, we evaluate the overlap between BASS and the 8-year \emph{Fermi}-LAT catalog (4FGL).

Sources in 4FGL are searched for around the position of each BASS AGN within a 0.2$^{\circ}$ radius. This is a conservative estimate of the position uncertainty of 4FGL sources as more than 96\% have better localizations (at 95\% CL). Of the 836 BASS sources, 84 have a 4FGL source within a 0.2$^{\circ}$ distance (10\%) and 61 (7\%) are consistent with the 4FGL position taking into account its localization uncertainty. As expected, the overlap (illustrated in Fig.~\ref{fig:zlumi}) increases for high-luminosity blazars.

\begin{figure}
    \centering
    \includegraphics[scale=0.5]{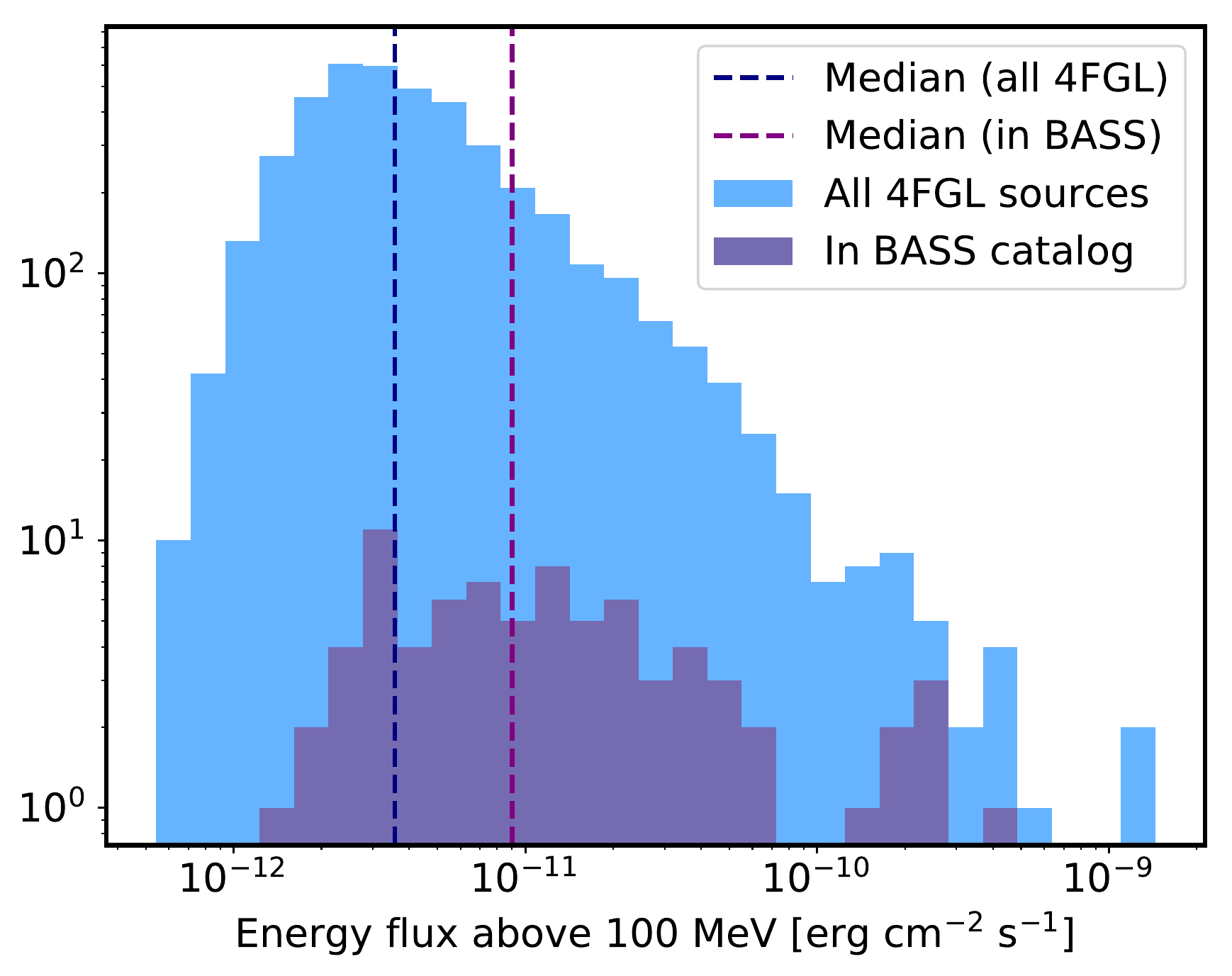}
    \caption{Gamma-ray energy fluxes above 100 MeV for all 4FGL sources (light blue) and those included in BASS (purple). The BASS gamma-ray sources have a slightly higher energy flux.}
    \label{fig:fermi_bass}
\end{figure}

To evaluate the overlap of the proposed study with those neutrino-\emph{Fermi} AGN correlation studies that use a gamma-ray flux weighting, we compare the distribution of energy fluxes for 4FGL sources included in BASS to the entire catalog of high Galactic latitude sources ($|b| > 5^{\circ}$). The 4FGL sources in BASS have a slightly higher median energy flux than the catalog median (see Fig.\ref{fig:fermi_bass}) but given their small overall representation in the sample (7\%) we estimate that the overlap with previous studies is not significant and therefore the BASS sample represents a novel source class selection to be evaluated as potential neutrino emitters.

The search for neutrino emission from the BASS AGN will use a time-integrated method that stacks neutrino candidate events from the positions of the BASS sources searching for an excess of events when compared to a background expectation. As the relation between the potential neutrino emission and the hard X-ray observables is not straightforward, different weighting schemes will be used, as well as a scheme that weights all sources equally regardless of their X-ray emission. 

\section{Data Sample and Analysis Method}

\subsection{Detector and data sample}

IceCube is a cubic-kilometer neutrino detector deployed deep within the glacier at the South Pole~\cite{Aartsen:2016nxy}. It detects neutrinos using a volumetric array of 5160 photomultiplier tubes that record the Cherenkov light emitted by relativistic charged particles produced in high-energy neutrino interactions in the ice or the bedrock underneath the detector. The energy, incoming direction, and flavor of the original neutrino can be inferred from the optical module signals. 

Charged-current interactions of muon neutrinos and antineutrinos lead to the production of energetic muons that can propagate over kilometers. The incoming direction of these muon tracks can be reconstructed to within 1$^{\circ}$ of the original neutrino at energies above 10 TeV. In this search we plan to use muon neutrino candidate events from the entire sky collected in 8 years of IceCube operating in its final configuration.

\subsection{Analysis method}

The search for neutrino emission will be performed using a time-integrated unbinned maximum-likelihood approach. The likelihood function is of the form

\begin{equation}
    \mathcal{L}(n_s, \gamma) = \prod_{i=1}^{N} \left[ \frac{n_s}{N} \mathcal{S}(\vec{x}_i, \gamma) + \left( 1-\frac{n_s}{N} \right) {\mathcal B}(\vec{x}_i) \right],
\label{eq:lh}
\end{equation}

where $n_s$ is the number of signal events, $N$ is the total number of events, $\vec{x_i}$ is a vector that contains spectral and positional information for each event, and $\mathcal{S}$ and $\mathcal{B}$ represent the source and background probability distributions, respectively. The parameter $\gamma$ is the spectral index of the assumed neutrino energy spectrum for all sources, modeled as a power law of the form $F_0 (E/E_0)^{-\gamma}$. The use of neutrino energy information can help further distinguish an astrophysical signal, which is expected to have a harder spectrum than the background ($\gamma_{\mathrm{bkg}} \sim 3.7$ for atmoshpheric neutrinos).

In a stacked search, the combined contribution from all sources to the likelihood function shown in Eq.~\ref{eq:lh} will be evaluated at once, with each AGN entering the source term with a weight assigned according to their observed properties $\omega^k$ and the sensitivity of the detector given the source declination $\delta_k$ and global spectral index $\gamma$. The source PDF for the entire catalog of $M = 828$ sources can then be written as 

\begin{equation}
\mathcal{S}_i = \frac{\sum_{k=1}^M \omega^k R^k(\delta_k, \gamma) \mathcal{S}(\vec{x}_i, \vec{x}_k, \gamma)}
{\sum^M_{k=1} \omega^k R^k(\delta_k, \gamma)}.
\label{eq:signalpdf}
\end{equation}

A test statistic $\Lambda$ is constructed from the likelihood ratio

\begin{equation}
    \Lambda = 2 \log \left[ \frac{\mathcal{L}(\hat{n_s}, \hat{\gamma})}{\mathcal{L}(n_s = 0)} \right]
    \label{eq:lhratio}
\end{equation}

where $(\hat{n_s}, \hat{\gamma})$ are the number of signal events and the spectral index for which the likelihood ratio with respect to the background-only hypotheses ($n_s = 0$) is maximized. The significance of any neutrino signal inferred from this analysis will be determined by comparing the $\Lambda$ value obtained by applying this method to the actual neutrino data to a high-statistics distribution of $\Lambda$ values obtained by scrambling multiple times the coordinates of the neutrino data set. 

\subsection{Proposed weighting schemes}

The weights $\omega_k$ assigned to each of the 828 BASS sources will be determined using five schemes: 

\begin{enumerate}
  \setlength\itemsep{0.em}
    \item \textbf{Flux weighting}: $\omega_k$ is proportional to the intrinsic flux of the AGN in the 14-195 keV band, which selects for near, higher flux sources.
    \item \textbf{Luminosity weighting}: $\omega_k$ is proportional to the isotropic-equivalent intrinsic luminosity of the AGN in the 14-195 keV band, which selects for more powerful objects.  The luminosity is calculated using a cosmological model with $\Omega_{\Lambda} = 0.7$, $\Omega_{M} = 0.3$ and $H_0$ = 70 km s$^{-1}$ Mpc$^{-1}$.
    \item \textbf{Spectral index weighting}: $\omega_k$ is inversely proportional to the X-ray power-law spectral index ($E^{-\Gamma}$). This selects for harder sources that may have significant flux at higher energies, specially towards the MeV range.
    \item \textbf{Column density weighting}: $\omega_k$ is proportional to the total column density of the source $n_{H}$, which selects for obscured sources with substantial target material along the line of sight.
    \item \textbf{Equal weighting}: $\omega_k$ is constant across the 
    catalog as an unbiased weight.
\end{enumerate}

\begin{figure}
    \centering
    \includegraphics[scale=0.4]{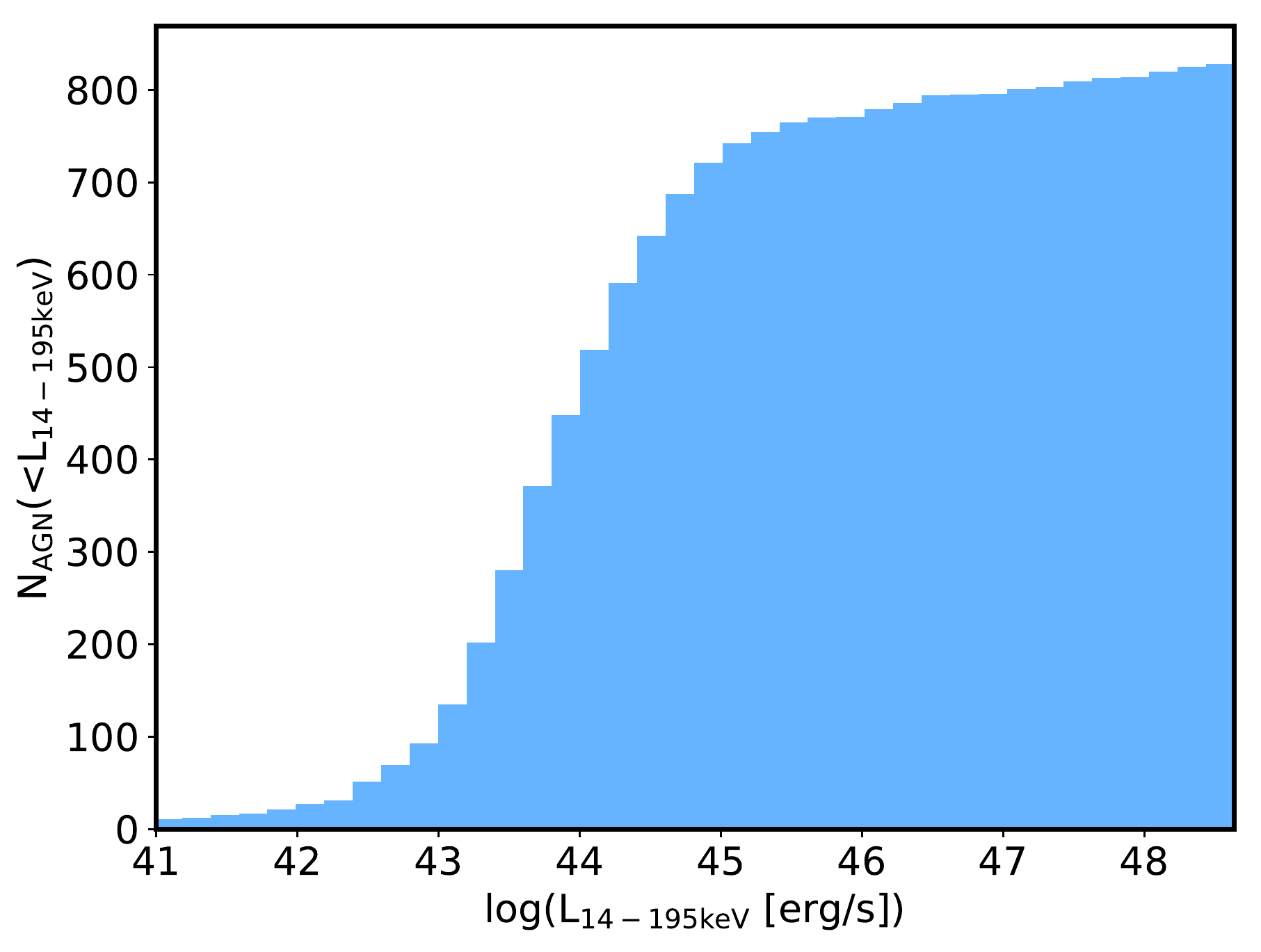}
    \includegraphics[scale=0.4]{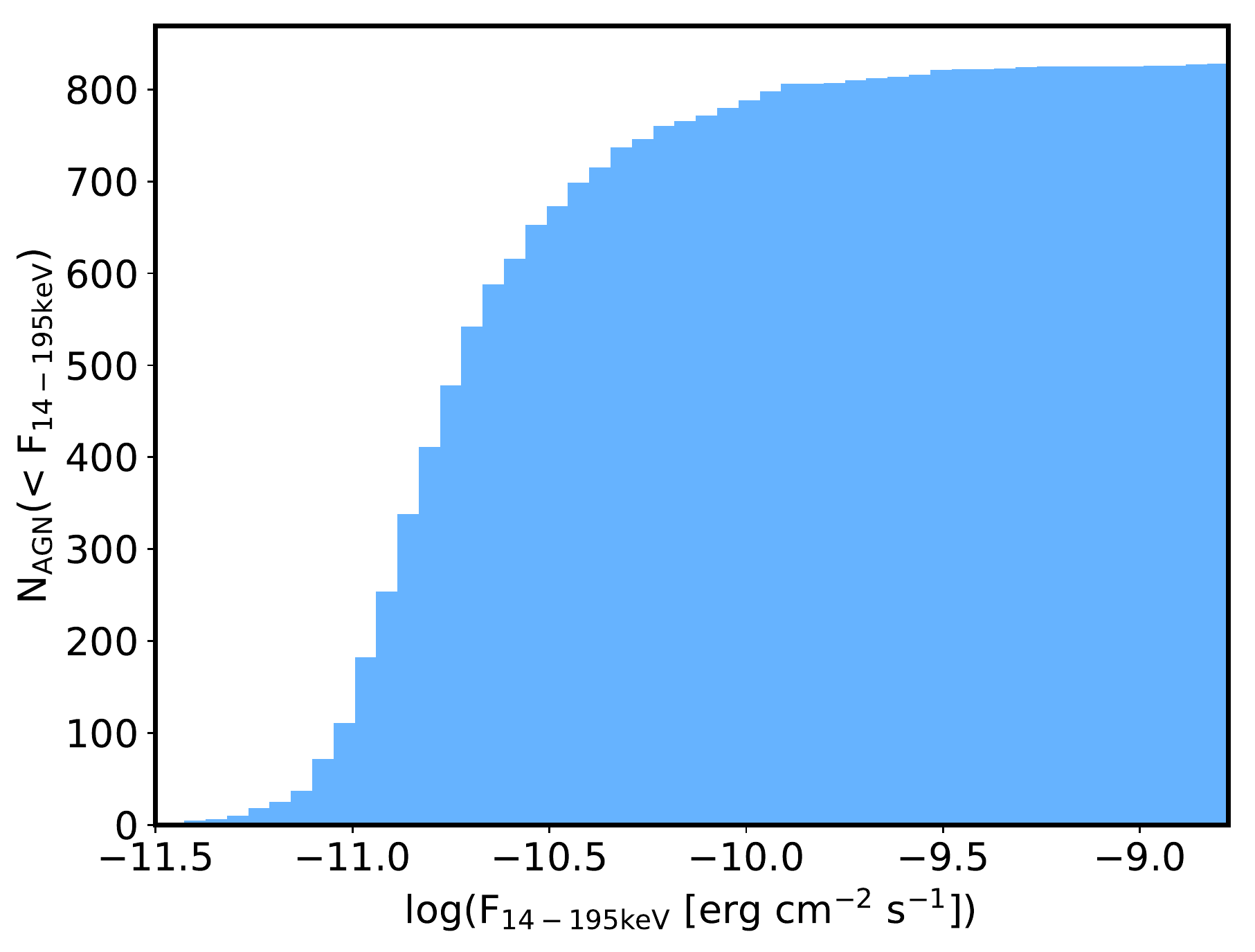}
    \includegraphics[scale=0.39]{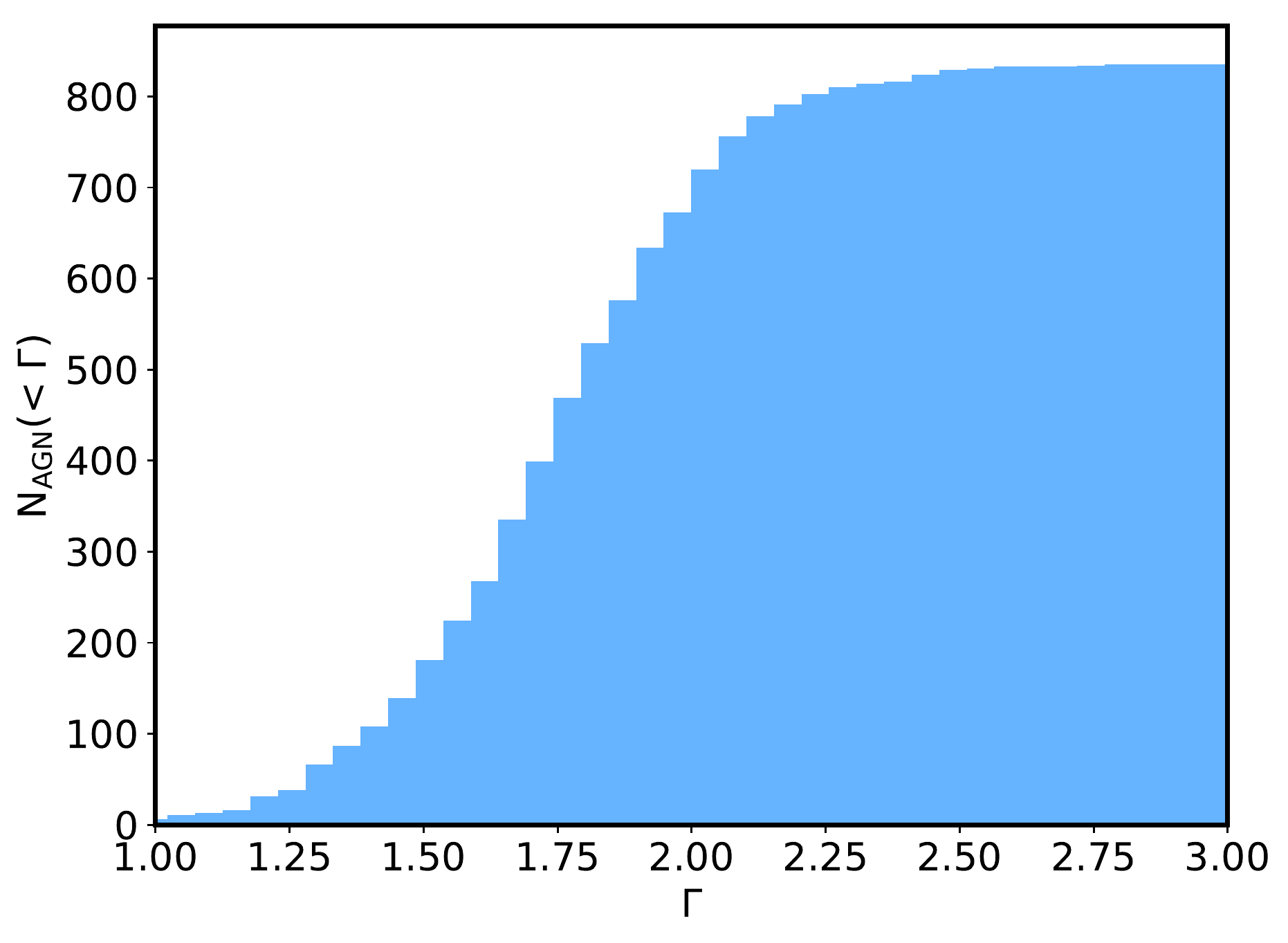}
    \includegraphics[scale=0.39]{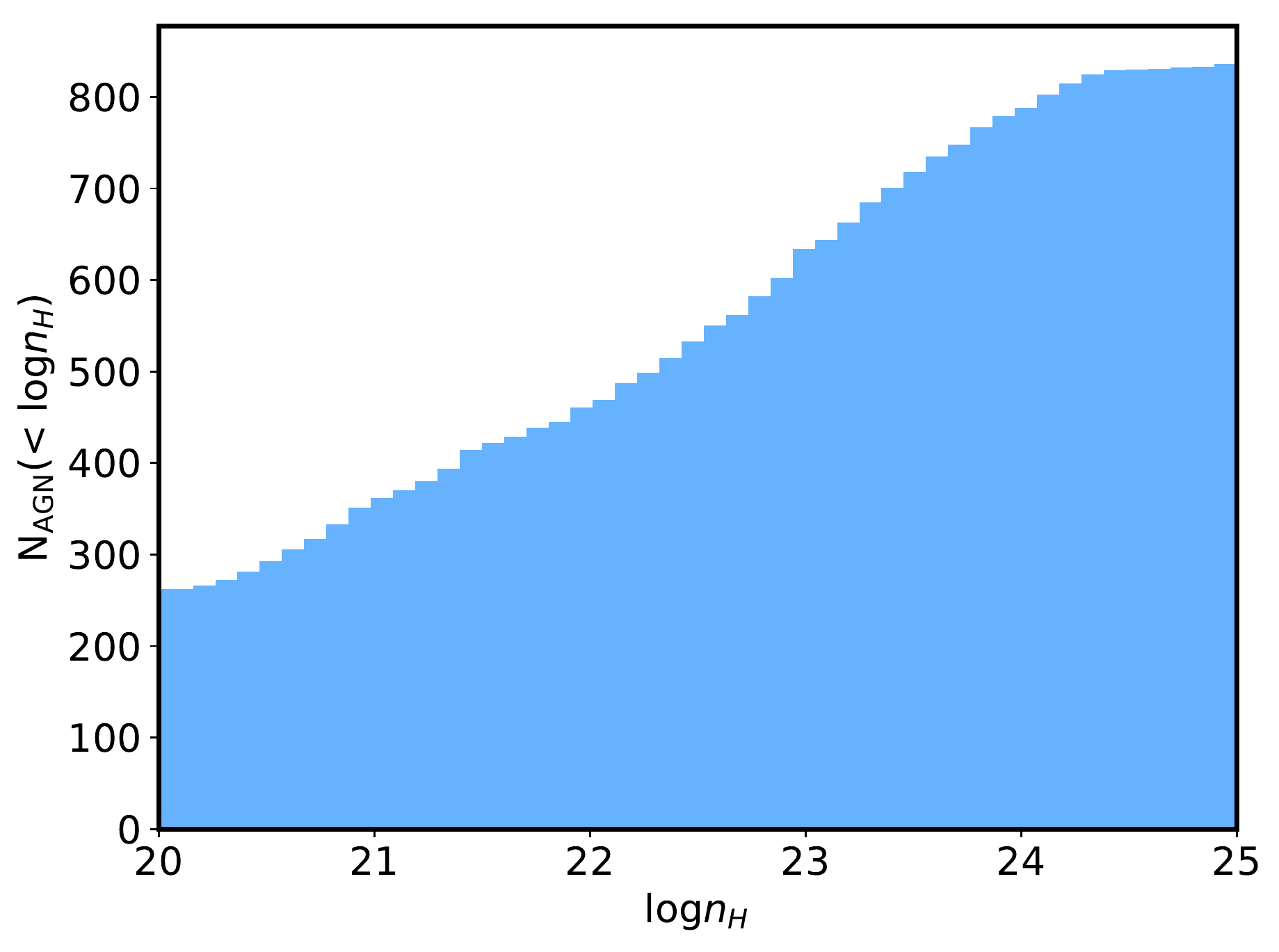}
    \caption{Luminosity (top left), intrinsic flux (top right), spectral index (bottom left) and column density (bottom right) weights based on the BASS catalog values.}
    \label{fig:weights}
\end{figure}

 The distributions of flux, luminosity, spectral index and column density weights are shown in Fig.~\ref{fig:weights}.

\section{Conclusions and future work}
 
We have discussed a proposed search for neutrino emission from hard X-ray AGN using 8 years of IceCube data and enumerated a weighting strategy for this analysis. Next steps include the generation of sensitivities and discovery potentials for this analysis and, in case no significant signal is identified, estimates of the constraints that can be set on this class of objects as contributors to the all-sky astrophysical neutrino flux.

%

\bibliographystyle{ICRC}
\bibliography{main}

\providecommand{\href}[2]{#2}\begingroup\raggedright\begin{thebibliography}{10}

\bibitem{Kelner:2006tc}
S.~R. Kelner, F.~A. Aharonian, and V.~V. Bugayov, {\em Phys. Rev.} {\bf D74}
  (2006) 034018. [Erratum: Phys. Rev.D79,039901(2009)].

\bibitem{Kelner:2008ke}
S.~R. Kelner and F.~A. Aharonian, {\em Phys. Rev.} {\bf D78} (2008) 034013.
  [Erratum: Phys. Rev.D82,099901(2010)].

\bibitem{2016RSOS....350555C}
A.~{Cooray}, {\em Royal Society Open Science} {\bf 3} (Mar, 2016) 150555.

\bibitem{Petropoulou:2015upa}
M.~Petropoulou, S.~Dimitrakoudis, P.~Padovani, A.~Mastichiadis, and E.~Resconi,
  {\em Mon. Not. Roy. Astron. Soc.} {\bf 448} (2015) 2412--2429.

\bibitem{PetroMast2015}
M.~{Petropoulou} and A.~{Mastichiadis}, {\em "Mon. Not. Roy. Astron. Soc."}
  {\bf 447} (Feb, 2015) 36--48.

\bibitem{Murase:2018iyl}
K.~Murase, F.~Oikonomou, and M.~Petropoulou, {\em Astrophys. J.} {\bf 865}
  (2018) 124.

\bibitem{Gao:2018mnu}
S.~Gao, A.~Fedynitch, W.~Winter, and M.~Pohl, {\em Nat. Astron.} {\bf 3} (2019)
  88--92.

\bibitem{Aartsen:2013jdh}
{\bf IceCube} Collaboration, M.~G. Aartsen et~al., {\em Science} {\bf 342}
  (2013) 1242856.

\bibitem{Aartsen:2016lir}
{\bf IceCube} Collaboration, M.~G. Aartsen et~al., {\em Astrophys. J.} {\bf
  835} (2017) 45.

\bibitem{Huber}
{\bf IceCube} Collaboration, M.~Huber, {\em These proceedings} {\bf 916}
  (2019).

\bibitem{IceCube:2018dnn}
{\bf IceCube, Fermi-LAT, MAGIC, AGILE, ASAS-SN, HAWC, H.E.S.S., INTEGRAL,
  Kanata, Kiso, Kapteyn, Liverpool Telescope, Subaru, Swift NuSTAR, VERITAS,
  VLA/17B-403} Collaboration, M.~G. Aartsen et~al., {\em Science} {\bf 361}
  (2018) eaat1378.

\bibitem{IceCube:2018cha}
{\bf IceCube} Collaboration, M.~G. Aartsen et~al., {\em Science} {\bf 361}
  (2018) 147--151.

\bibitem{Rodrigues:2018tku}
X.~Rodrigues, S.~Gao, A.~Fedynitch, A.~Palladino, and W.~Winter, {\em
  Astrophys. J.} {\bf 874} (2019) L29.

\bibitem{Barthelmy:2005hs}
S.~D. Barthelmy et~al., {\em Space Sci. Rev.} {\bf 120} (2005) 143.

\bibitem{Koss:2017ilk}
M.~Koss et~al., {\em Astrophys. J.} {\bf 850} (2017) 74.

\bibitem{Aartsen:2016nxy}
{\bf IceCube} Collaboration, M.~G. Aartsen et~al., {\em JINST} {\bf 12} (2017)
  P03012.

\end{thebibliography}\endgroup

\end{document}